# Optical Communication on CubeSats
## – Enabling the Next Era in Space Science –

Alberto Carrasco-Casado
Space Communications Laboratory
NICT
Tokyo, Japan
alberto@nict.go.jp

Abhijit Biswas
Jet Propulsion Laboratory
NASA
Pasadena, USA
abhijit.biswas@jpl.nasa.gov

Renny Fields
The Aerospace Corporation
El Segundo, USA
renny.a.fields@aero.org

Brian Grefenstette
Space Radiation Laboratory
California Institute of Technology
Pasadena, USA
bwgref@srl.caltech.edu

Fiona Harrison
Cahill Center for Astrophysics
California Institute of Technology
Pasadena, USA
fiona@srl.caltech.edu

Suzana Sburlan
Amazon
Los Angeles, USA
suzana.sburlan@gmail.com

Morio Toyoshima
Space Communications Laboratory
NICT
Tokyo, Japan
morio@nict.go.jp

*Abstract*—CubeSats are excellent platforms to rapidly perform simple space experiments. Several hundreds of CubeSats have already been successfully launched in the past few years and the number of announced launches grows every year. These platforms provide an easy access to space for universities and organizations which otherwise could not afford it. However, these spacecraft still rely on RF communications, where the spectrum is already crowded and cannot support the growing demand for data transmission to the ground. Lasercom holds the promise to be the solution to this problem, with a potential improvement of several orders of magnitude in the transmission capacity, while keeping a low size, weight and power. Between 2016 and 2017, The Keck Institute for Space Studies (KISS), a joint institute of the California Institute of Technology and the Jet Propulsion Laboratory, brought together a group of space scientists and lasercom engineers to address the current challenges that this technology faces, in order to enable it to compete with RF and eventually replace it when high-data rate is needed. After two one-week workshops, the working group started developing a report addressing three study cases: low Earth orbit, crosslinks and deep space. This paper presents the main points and conclusions of these KISS workshops.

## I. INTRODUCTION

CubeSats are miniaturized satellites built in increments of 10×10×10 cm cubes: one 10-cm$^3$ cube is called 1U (one unit), two cubes together are called 2U, etc. Although CubeSats are classified according to their size, conventional satellites are usually classified by their deployed mass (see Table 1) because it has a more direct relation with their launching cost. In this regard, CubeSats could be classified as picosatellites, nanosatellites or microsatellites, being nanosatellites the most frequent type of CubeSat since around 80% of them have formats between 1U and 3U [1], and as a rule of thumb, each 1U is usually associated with 1 kg of mass. The most popular CubeSat form factor is 3U, followed by 1U, 2U, and 6U.

Table 1. Classification of satellites according to their mass [2].

| Satellite type | Mass (kg) |
|---|---|
| Large satellite | > 1000 |
| Medium satellite | 500 to 1000 |
| Mini satellite | 100 to 500 |
| Micro satellite | 10 to 100 |
| Nano satellite | 1 to 10 |
| Pico satellite | 0.1 to 1 |
| Femto satellite | < 0.1 |

The CubeSat reference design, which became a de-facto standard, was originally proposed in 1999 [3] and only defines the exterior form factor and a simple deployer with a capacity of 3U (although later new designs were proposed with formats bigger than 3U [4]). This was conceived with a main point in mind, i.e. protect the launch vehicle, which proved to be a key point to the rapid increase in launch rate of these spacecraft. The goal of the original design was to enable graduate students to design, build, and fly their own designs before they graduated, avoiding the traditionally long time of development, too high cost, and few available launches associated to conventional satellites. Almost 800 nanosatellites (see Fig. 1) have already been successfully launched in the past few years and the number of announced launches grows every year.

CubeSats often use COTS (Commercial Off-The-Shelf) components with very relaxed radiation requirements, which makes them excellent platforms to rapidly perform simple space experiments and provide an easy access to space for universities and organizations which otherwise could not afford it. CubeSats can be accommodated as secondary payload in any launch vehicle, or be delivered into space as cargo to be deployed from the ISS (International Space Station).

While CubeSats used to be thought as platforms designed to carry out simple tests in space in a cost-effective way for teaching and demonstration purposes, the technology is reaching such a level of maturity that it allows to think of CubeSats in a radically different way: they are proving to be an important scientific tool, and can potentially address relevant goals in the fields of space and Earth sciences. There are already a number of examples where CubeSats have provided valuable scientific results, published in top journals [5] [6], achieved with a relatively-small amount of invested money, and many new mission concepts have been identified where CubeSats could play an important scientific role in fields like astrophysics, heliophysics, geophysics and planetary exploration. High data-rate communication has been identified as one of the key enabling technologies to carry out many of these mission concepts [7].

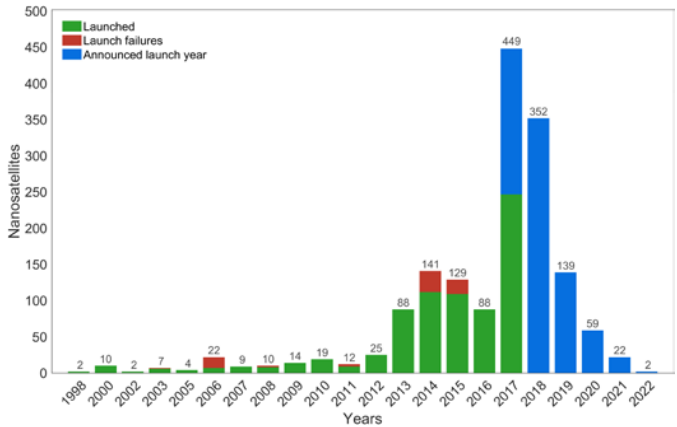

Fig. 1. Nanosatellites by their announced launch year [1].

## II. KISS: KECK INSTITUTE FOR SPACE STUDIES

The Keck Institute for Space Studies (KISS) is a joint institute of the California Institute of Technology (Caltech) and the Jet Propulsion Laboratory (JPL), founded in January 2008 with a grant of 24 million dollars from the W. M. Keck Foundation. This privately-funded think tank is listed in the "2016 Global Go To Think Tank Index Report" [8] at the 28th position in the "Top Science and Technology Think Tanks" category. Its primary purpose is to bring together a broad spectrum of scientists and engineers for sustained technical interaction aimed at developing new ideas for revolutionary advances in space missions. Once a key innovation for a new mission concept is identified, the Institute funds the initial steps towards making progress on that challenge.

52 programs were carried out between 2008 and 2016, and other 6 are ongoing in 2017, with over 200 lectures and more than 40 studies already completed, openly available through the KISS website [9]. KISS studies are carried out in the form of one or more 1-week full-time workshops, where the invitation-only participants actively contribute in brainstorming sessions, lightning talks, and group/subgroup discussions, after which the working groups come up with a study report describing the main conclusions reached during these workshops.

The study called "Optical Communication on smallsats. Enabling the Next Era in Space Science" was led by Abhijit Biswas (JPL-NASA), Renny Fields (The Aerospace Corporation), Brian Grefenstette (Caltech), Fiona Harrison (Caltech) and Suzana Sburlan (JPL-NASA at the time of the workshop, currently Amazon). It comprised two workshops: The first one was held on July 11-14, 2016 and the second one on February 1-9, 2017, both at the KISS facilities, in Caltech (Pasadena, California). The goal of this KISS study was to identify the most promising development paths for optical communications that will enable CubeSat missions operating from near-Earth to deep-space, in order to overcome the key risks associated with this technology and ensure that it will meet the needs of CubeSat customers and be competitive with its RF counterpart.

## III. COMMUNICATION SYSTEMS IN CUBESATS

Currently, CubeSats rely completely on radiofrequency communications (see Fig. 2). Most of the University-class CubeSats use amateur low-speed UHF systems (with omnidirectional dipole antennas) due to its availability and lower cost, with rates in the order of kbit/s or tens of kbit/s from Low-Earth Orbit (LEO) [10], and only reaching Mbit/s with 20-m class ground antennas and special government bandwidth allocation [11]. As a result, the potential of most CubeSat missions is being limited by their communication capabilities, although mainly due to regulatory issues rather than to technological limitations. The RF spectrum is already very crowded, especially in these lower parts of the spectrum, and getting an allocation from ITU is usually the hardest part of a typical CubeSat mission, taking even longer times than the CubeSat development itself, thus sometimes risking the launch opportunities [12]. With a foreseen increase in the number of CubeSat missions, the current RF spectrum will not be able to support the growing demand for data transmission to the ground. Besides, this congested part of the spectrum presents a higher risk of interference with other systems.

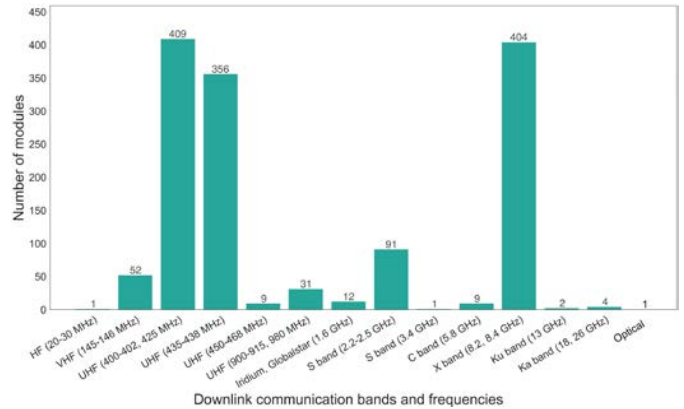

Fig. 2. Nanosatellite downlink communication bands [1].

As a solution to the problems explained before, there is a tendency to move to higher frequencies, especially to X-band [13], where more bandwidth is available, as an alternative to achieve higher data rates with smaller ground antennas. Although these transmitters started being developed several years ago [14] [15], and they have begun to be commercially available, they present their own challenges, including the relatively higher cost, the higher energy consumption, and the challenging pointing requirements for the directional antennas.

The Earth-imaging company Planet, currently with 120 operational 3U CubeSats, has shown a remarkable success in these X-band communication systems [16], using COTS components and 5-m class ground antennas to achieve sustained data rates in the order of 100-200 Mbit/s with the latest generation of their 'Dove' CubeSats. Forthcoming improvements in these X-band systems are expected with more effective power generation, better pointing accuracy and increased antenna gain.

## IV. LASERCOM WITH CUBESATS

A survey of the 49 CubeSats launched between 2009 and 2012 concluded that the communication system is a major limiting factor for CubeSat development [17]. Using optical wavelengths is the next logical step in the tendency of CubeSats communications (as well as satellite communications) migrating to shorter and shorter wavelengths. Laser communications (lasercom) holds the promise to be a solution to this problem, with a potential improvement of several orders of magnitude in transmission capacity, while keeping a low size, weight and power. However, as shown in Fig. 2, so far there has not been any successful demonstration of this technology in a CubeSat. Considering only lasercom terminals with a mass of less than 10 kg, only two missions have successfully flown this type of systems: NICT's SOTA (Small Optical TrAnsponder) and DLR's OSIRIS (Optical Space Infrared Downlink System). These systems are still not suitable for CubeSat platforms, but they have been the closest successful attempts in this direction.

### A. History of lasercom terminals in small satellites

An impressive first test of basic optical communication onboard a 1U CubeSat was performed by the Fukuoka Institute of Technology (Japan) with FITSAT-1, also known as Niwaka, which was deployed by the robotic arm of the International Space Station on October 2012. Niwaka had a neodymium magnet as a passive attitude control system that made the top panel face the Fukuoka ground station. This panel contained 50 green 3W LEDs, achieving 200-W pulses, modulated with a 1-kHz Morse-code signal. These signals were received using a photomultiplier coupled to a 25-cm ground telescope [18].

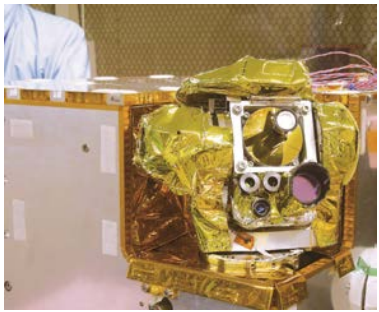

Fig. 3. NICT's SOTA flight model onboard SOCRATES.

The first proper lasercom system onboard a small satellite was NICT's SOTA [19] onboard SOCRATES (Fig. 3), which was launched in May 2014 into a ~600 km LEO orbit, and it was fully operative until November 2016. SOTA was a 2-axis gimballed terminal with capabilities to perform a variety of lasercom experiments in a less-than-6 kg compact package. The core experiment was the 10 Mbit/s links at 1549 nm. This capability used the main subsystems of SOTA, i.e. a coarse-pointing-assembly to track the NICT OGS at Koganei (Tokyo, Japan), a receiving lens to track the OGS 1-µm beacon laser using a Silicon quadrant detector, a fine-pointing-assembly to accurately transmit the 35-mW laser through a 5-cm Cassegrain telescope, and an electronics unit to generate the stream of data, codify it using error correcting codes, and interleave the bits against signal fading. SOTA had other additional capabilities, i.e. B92-like QKD protocol using two 800-nm band lasers with linear polarizations separated by 45° to perform a quantum-limited basic demonstration [20], and 10 Mbit/s downlinks at 980-nm using a lens, both based on coarse-pointing only.

Table 2. Specs. of lasercom terminals onboard small satellites.

|  | **SOTA** | **OSIRISv2** |
|---|---|---|
| Operator | NICT, Japan | DLR, Germany |
| Launch date | May 24, 2014 | June 22, 2016 |
| Satellite | SOCRATES (48 kg) | BIROS (130 kg) |
| Mass | 5.9 kg | 5 kg |
| Size | 18×11×10 cm | 25×20×10 cm |
| Beacon | 1 µm unmodulated | 1560 nm modulated |
| Downlink | 800, 980, 1549 nm | 1545, 1550 nm |
| Modulation | On-Off Keying | On-Off Keying |
| Max. bitrate | 10 Mbit/s | 1 Gbit/s |

DLR's OSIRIS onboard BIROS, known as OSIRISv2 [21], was the second, and so far, the latest, lasercom terminal onboard a small satellite. BIROS was launched in June 2016 into a 500-km LEO orbit. OSIRISv2 has almost the same mass as SOTA, but unlike SOTA, it does not include a coarse-pointing system. Instead, a closed-loop satellite body pointing acts as the coarse control, tracking the 1560-nm modulated beacon with an InGaAs quadrant-type tracking sensor. The main demonstration consists in downlinks of up to 1-Gbit/s using an OOK-modulated 1-W 1545-nm laser through a 1.5-cm lens with a 200-µrad divergence. OSIRISv2 also includes another downlink capability up to 150 Mbit/s using a separate 1-5-cm lens with a divergence of 1200 µrad, a transmitted power of 150-mW at 1550 nm, and open-loop control based on the satellite attitude instead of the beacon tracking sensor.

### B. Principles of lasercom with CubeSats

The key components of a CubeSat lasercom system are the optical-power generation and the pointing capability. Like most of the other spacecraft, CubeSats generate power out of photovoltaic cells, either body-mounted or on deployable panels. In this regard, the size of CubeSats makes power generation the most obvious limitation. The efficiency of solar cells is increasing from the 20% of single junction to the typical 30% of triple junction cells, usually used in CubeSats, although ongoing research on multi-junction cells is expected to increase it to get close to the theoretical limit of 86.6% [22]. Assuming a typical 28% efficiency, depending on the geometry of the solar panels and the sun exposure, a LEO 3U CubeSat can generate an average power in the order of 5-10 W and peak power of ~20 W [23].

Generally speaking, laser sources suitable to be mounted on CubeSats can be divided in two categories. The most straightforward strategy is using directly-modulated high-power laser diodes at wavelengths in the 900–1000 nm range. Their electrical-to-optical efficiency exceeds 50% [24] and they show a good size/weight balance, being the bandwidth their main limitation, in the order of 100 MHz. For higher bandwidths, the best alternative is the Master Oscillator Power Amplifier (MOPA) architecture. This technology is available either at 1064 nm or 1550 nm, and it is based on fiber amplification as their key element, delivering an optical power in the order of several W. Since the main limitation is the average power, this technology also makes it possible to implement a dynamic adaptation to the channel conditions by means of the Pulse-Position Modulation (PPM). Using PPM, the number of bits per pulse grows with the number of modulation symbols M, and keeping the same average power, it is possible to reduce the duty cycle (1/M), transmitting more photons per bit with higher peak power at a lower data rate [25]. Generating several Watts of optical power imposes requirements that go beyond what a CubeSat can usually deliver, although an effective strategy to alleviate this is to include secondary batteries in the power management system.

Any lasercom system onboard a CubeSat needs some kind of coarse pointing. The attitude determination and control system can offer this capability instead of a gimbal, which is usually too heavy. This system usually consists of a star tracker and a magnetometer for determination, and reaction wheels and magnetorquers for control. An accuracy in the order of 0.01° (3σ) has been demonstrated, and it is already commercially available and with flight experience [26]. This enabling technology makes it possible to relax the requirements of the transmitted power, and/or the additional fine pointing system. In the first case, if the CubeSat is based solely on body-pointing and this is very accurate, a narrower beam can be transmitted, maximizing the energy transfer. In the second case, when additional fine pointing is required, an accurate body pointing enables using a better resolution, since there is a dependence between the total range and the maximum resolution (this ratio is in the order of 1:10,000 [27]), i.e. a smaller range makes it possible to achieve a better resolution.

With the lower beam divergence that laser sources and pointing accuracy allow, a new requirement needs to be taken into consideration, i.e. the point-ahead angle. Due to the finite speed of light, it takes some time for a transmitted beam to reach the receiver. Therefore, the transmitted and received beams are angularly separated by the so-called point-ahead angle. This should be included in the fine-pointing system applying an angular shift to the transmitted beam if the point-ahead angle is comparable or larger than the beam width. As a reference, the point-ahead angle with the Earth is 51 μrad for 500-km LEO, 20 μrad for GEO, and up to 387 μrad in Mars.

Regarding the ground segment, although frequently the complexity is shifted from the satellite side to the ground in order to reduce cost, this is not usually the case with CubeSats due to the low cost associated with these missions. Furthermore, the requirements for the optical ground stations are not especially demanding, and they can be met with commercial telescopes and COTS components. The straightaway solution for the receiving part consists of an Avalanche Photodiode operated in the linear mode, typically with large active area which allows free-space or multimode-fiber coupling (minimizing the losses due to atmospheric turbulence) of up to Gbit/s rates, followed by a transimpedance amplification stage. This configuration allows achieving sensitivities of several hundreds of photons/bit. More sophisticated architectures have shown performances well below 100 photons/bit, but with strong requirements such as adaptive optics + single-mode fiber coupling + low noise pre-amplification or superconductive nanowire photodetector arrays, both options involving a complexity far beyond the scope of usual CubeSat missions.

*C. Low-Earth Orbit (LEO) links*

Since small satellites are usually located in LEO [28], this appears to be the most obvious scenario for CubeSats as well, which have not operated beyond LEO yet. This orbit is very suitable because of the more benign radiation environment, the close distance to Earth, and the more frequent launches. Furthermore, their lower cost makes them suitable not only for single experiments in low orbit, but for deployment in large LEO constellations as well, where they can provide a unique coverage of the entire Earth. Orbits beyond LEO make it very difficult to fulfill the 25-year post-mission lifetime guideline set by the IADC (Inter-Agency Space Debris Coordination Committee) in order not to become space debris. For example, CubeSats in orbits above 750 km take centuries to decay [29].

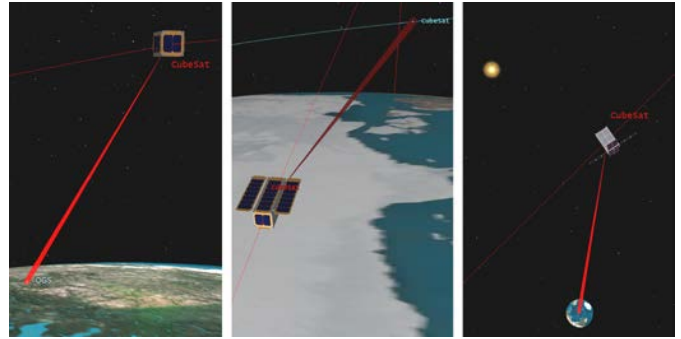

Fig. 4. Scenarios for potential CubeSat missions. Left: Low-Earth Orbit, center: crosslinks, right: deep space.

The main goal of this scenario is direct LEO-to-ground downlinks (see Fig. 4, left-hand side), since the main point is being able to download the large amount of data that the increasing number of sensors onboard CubeSats require. The operation of a typical pass in this scenario is similar to other lasercom LEO missions: before the scheduled pass, an RF link is used to communicate with the satellite updating the orbital data and other relevant pass information, and when the satellite is within the line of sight, it rotates facing the tracking sensor towards the ground station, and a powerful beacon is transmitted from the OGS, being used by the satellite as a reference to close the tracking loop of the body-pointing and the fine pointing system, in case there is one, until the beacon is lost or the communication gets degraded.

The LEO scenario usually implies infrequent and short passes over a given ground station (except the rare cases when the ground station is located in the poles). The link frequency and duration strongly depends on the maximum link range for a given orbit altitude, which is determined by the minimum ground station elevation. The Fig. 5 shows the dependence of the average link duration and frequency, and maximum distance with the ground station elevation for three LEO heights. This result is calculated for the NICT's OGS at Koganei (Tokyo, Japan), which is located in an average latitude (35°41'58'') in the Northern Hemisphere. As for the chosen LEO heights, 300 km is a very-low case, considering that below 200 km a CubeSat will decay within one day [30], and 700 km is a very-high case considering that above 750 km CubeSats take centuries to decay [29]. For an average 500-km orbit (Fig. 5, yellow lines), the maximum link duration would be ~7 minutes for a 5° elevation, which implies a ~2,000 km distance. In practice, a simple design usually requires to find a compromise between link distance and link duration, such as 20° of elevation and ~1,000 km of link distance. This assumption implies ~4-minutes links and ~2 links per day. This calculation considers night and day communications, which is an optimistic assumption for a CubeSat mission. In practice, if only nighttime links are allowed, the frequency would be even smaller. The solution to this problem is site diversity or the use of onboard memory to storage the data to be transmitted.

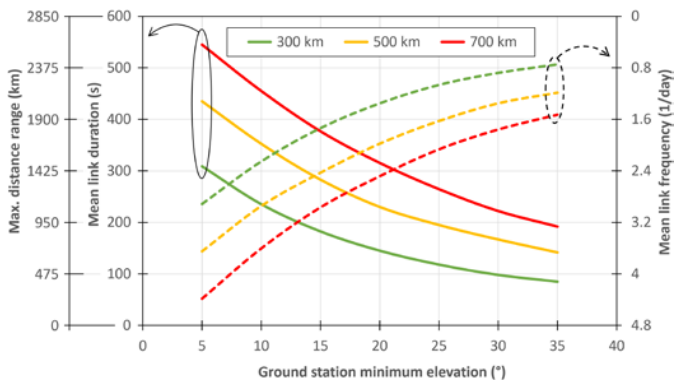

Fig. 5. Dependence of the link duration, frequency and distance range with the ground station elevation for three LEO heights at NICT's OGS at Koganei (Tokyo, Japan).

In terms of laser communications technology, the main challenge in this scenario, with fast motion and short passes, is the pointing accuracy. Although early CubeSats had very limited or no attitude control, this is not the case any more, and as this technology evolves, it becomes more and more commercially available, enabling very directional communications, as lasercom requires. The omnidirectional or very wide RF beams can then be replaced by narrow laser beams, and in LEO this is possible even using very-small transmitting apertures. For example, a 1-km footprint can be obtained with a diffraction-limited aperture as small as ~1 mm, or a ~100-m footprint with a ~1-cm aperture.

If the pointing accuracy requires transmitting above ~1 W of average optical power, power amplification is generally required, which is more efficient in the 1064-nm wavelength (YDFA) compared to the 1550-nm alternative (EDFA): a "wall-plug" efficiency of ~25% vs ~10%, respectively. In this regard, a good pointing accuracy makes it possible to relax the transmitted-power requirement in the LEO scenario. Therefore, if power amplification is not required, 1550 nm is preferable due to the better availability of telecom components and the better atmospheric-propagation behavior.

A simple modulation scheme that adapts well to the LEO-to-ground scenario is the On-Off Keying (OOK). Although the quantum-limited sensitivity of this modulation is 33.9 photons/bit [31], the thermal noise makes practical implementations show much-worse sensitivities. 1,000 photons/bit is a conservative assumption for high-performance communications (Gbit/s class and better than $10^{-9}$ Bit Error Rates) [32]. With an available body-pointing accuracy in the order of 0.004° (see section IV.B), which allows using a diffraction-limited 1-cm lens to produce a 155-µrad beam divergence at 1550 nm, assuming a 500-mW transmitted power, a minimum OGS elevation of 20° (maximum link distance of 1200 km for a 500-km orbit), and a receiver sensitivity of 1,000 photons/bit, a 100-Mbit/s link could be closed with a 40-cm ground telescope.

The system described above is probably the most straightforward lasercom configuration for LEO CubeSats with state-of-the-art, yet COTS, components. Based on those assumptions, the bit rate could be increased up to the Gbit/s order by using a fine-pointing mechanism onboard the CubeSat to achieve ~20 µrad accuracy, being able to transmit a several-times narrower beam (~50 µrad) with a bigger, but still feasible, aperture (~3 cm). Alternatively, the same Gbit/s regime could be achieved by the first body-pointed system configuration, with no specific fine-pointing mechanism, and a higher transmitted power (~5 W). This option comes at the cost of bigger volume and mass requirements, due to the power amplifier and the necessary solar panels and/or secondary batteries.

*D. Intersatellite links*

Crosslinks, or intersatellite links, are defined as communication links that begin and end in space. Interconnecting spacecraft through crosslinks has a wide range of advantages and possible applications. When using CubeSats in one or both ends of the communication (see Fig. 4, center), four important scenarios can be portrayed: in the first one, deep-space probes connect with the Earth through an intermediate near-Earth node in a repeater architecture; the second scenario consists in connecting LEO with GEO, improving the availability of LEO satellites by using a GEO satellite as a repeater; a third scenario would connect multiple satellites within a constellation to facilitate the data download to the ground stations; and the fourth scenario connects spacecraft working in a mother-daughter architecture, for example, where one of the probes is intended to crash in proximity operations approaching asteroids. Other important applications beyond lasercom include power beaming to remote areas.

The four previous scenarios greatly differ in the distance range. In the first one (deep-space to near-Earth repeater), the distances exceed 300,000 km assuming the Moon as the limit for deep space; in the second scenario (LEO-to-GEO), the distance would be in the order of 40,000 km; the third scenario

(constellations) implies distances ranging from hundreds to thousands of km, assuming LEO constellations; and in the third scenario (mother-daughter architecture), the distances could be as short as several km. With several orders of magnitude of distance difference, there can be no single solution to such different scenarios. With current technology, the two latter scenarios could certainly be implemented. However, technological gaps are yet to be solved in order to implement crosslinks in the LEO-to-GEO and Deep-space to near-Earth scenarios to be able to comply with the severe size, weight and power requirements of CubeSats.

*E. Deep-Space links*

Although deep-space missions go beyond the original concept which CubeSats were conceived for, there are already a number of deep-space mission concepts based on CubeSats unfeasible with current RF communications where lasercom could be an enabling solution. Radio-frequency links become less and less efficient with the long distances of deep space due to the wide beam divergence. Optical wavelengths allow much narrower beams (see Fig. 4, right-hand side), optimizing the transmitted energy, although the maximum apertures, which determines the minimum divergence, are also constrained by the small CubeSat form factor.

Deep-space missions have very special requirements, which needs current CubeSat technology to evolve to fill several implementation gaps. One of them is the much longer mission duration: whereas LEO missions usually do not require to last longer than several months, deep-space missions usually require several years of survival in a harsher radiation environment, which forces using a higher grade of components compared to the ones usually used in LEO missions. Depending on the mission, the available power could be smaller due to the longer distance to the Sun, and accordingly the required allocation for solar panels and batteries would be bigger. Deep-space spacecraft need some kind of propulsion system as well in order to be able to reach their final destination. Also, as opposed to the LEO scenario, in deep space the transmitting aperture plays an important role, to enable reducing the divergence to deliver enough power to the Earth. For all these reasons, a more suitable CubeSat form factor for deep space is 6U or bigger. The requirements of ground stations are much more demanding as well, normally requiring apertures exceeding one meter and more sophisticated and sensitive receivers.

As a baseline design for deep space, based on a system like the one described for LEO (OOK modulation, 500-mW transmitted power, 0.004°-accuracy body pointing, 10-μrad fine-pointing accuracy, and 1,000-photons/bit receiver sensitivity), a 1-Mbit/s link could be closed from the Moon with a 1-m receiving telescope if the transmitted aperture was increased to almost the maximum size allowed in a CubeSat, i.e. about 8 cm to produce a diffraction-limited 20-μrad beam divergence. The same system could close a 100-kbit/s link from the Sun-Earth Lagrange point L2 at 1.5 million km, or roughly 50 bits/s from Mars in opposition (75 million km) and 2 bits/s from Mars in conjunction (375 million km). Since the available data rate from deep space is very limited and the speed constraints are not very demanding, these systems can be improved by using PPM, which allows transmitting a higher peak power with the same average power. For example, just by changing the modulation in the previous system from OOK to 16-PPM, the data rate of the Mars scenarios in opposition and conjunction would improve from 50 and 2 bits/s to 800 and 30 bits/s, respectively. Further improvements could come from the increase of the transmitting aperture (although limited by the CubeSat form factor), a higher transmitted power, a higher PPM-modulation order, a more sensitive receiver (with any of the strategies mentioned in the section IV.B), or the increase of the receiving aperture.

CONCLUSION

The CubeSat market is growing rapidly, reducing all the costs associated with completing a mission as well as the period from design concept to experiment, offering better technologies with better specifications, and more and cheaper launch opportunities. As in the other past technological revolutions, the best trigger for a technology to develop to its full potential comes with the access to a mass market. CubeSats have the capability of democratizing space technologies, making them accessible for the first time to a wide variety of users, ranging from universities to small companies. Space optical communications could play an important role, enhancing the potential of CubeSats with growing bandwidth requirements. The time will soon come when the lasercom option will be offered as a standard solution for high-speed communications.

As a scientific and technological think tank, the Keck Institute for Space Studies aims to anticipate potential breakthroughs and paradigm shifts, as well as actively facilitate its accomplishment. Free-space optical communication has been identified as an enabling technology for many new space mission concepts, but it needs further development to reach its full potential. To achieve these goals, the KISS working group recommends the scientific community to start planning missions that leverage the enhanced capabilities provided by lasercom; the space agencies and research institutes to support the development and demonstration of lasercom technologies; and the industry to anticipate the needs of future CubeSat users accelerating the commercialization and availability of these technologies.

ACKNOWLEDGMENT

This work was supported in part by the W. M. Keck Institute for Space Studies [33]. The authors also want to thank all the participants of the two workshops: Krisjani Angkasa (JPL), Alessandra Babuscia (JPL), Derek Barnes (MIT), Kerri Cahoy (MIT), Amir Caspi (SwRI), Emily Clements (MIT), Sam Dolinar (JPL), Peter Goorjian (NASA), Varoujan Gorjian (JPL), Brian Gunter (Georgia Institute of Technology), Frank Heine (Tesat Spacecom), Travis Imken (JPL), Farzana Khatri (MIT), Maxim Khatsenko (MIT), Ryan Kingsbury (Planet Labs), Jonathan Klamkin (UC Santa Barbara), David Klumpar (Montana State University), Michael Krainak (NASA), Michael Küeppers (ESA), Joe Kusters (JPL), Myron Lee (MIT), Rachel Morgan (MIT), Dhack Muthulingam (JPL), Michael Peng (JPL), Sean Pike (Caltech), Joseph Riedel (JPL), Kathleen Riesing (MIT), Bryan Robinson (MIT), Darren

Rowen (The Aerospace Corp.), Joel Shields (JPL), Harlan Spence (University of New Hampshire), Mark Storm (Fibertek Inc.), Jan Stupl (SGT/NASA), Jose Velazco (JPL) and Hua Xie (JPL).